# William H. Kruskal and the Development of Coordinate-Free Methods

**Morris L. Eaton**

Soon after joining the University of Chicago's Statistics Department, in the fall of 1966, I became aware of Bill Kruskal's lecture notes on topics he wryly referred to as "a coordinate-free approach to linear this and that." Graduate students raved about his course and the three-inch set of lecture notes that was clearly valued property. Bill's interest in applying vector space methods had been spurred by Jimmie Savage in the mid-1950s. Indeed to quote Bill, "Once Jimmie said a few magic words, it all became plain, but it needed writing up" (see Zabell, 1994, page 293). My guess is that the "writing up" had been going on for almost a decade prior to my arrival in Chicago.

By 1966 the notes were quite polished and consisted of nine or so chapters of linear statistical model theory that were used in a two-quarter course at Chicago. The care with which the notes were prepared, very characteristic of Bill, certainly suggested an intent to publish a book based on them. He had joined the Chicago faculty in 1950 and had finished his Ph.D. in 1955. It thus seems reasonable to conjecture that serious work on the notes was begun in the late 1950s (an early research paper in the area is Kruskal, 1961). In discussions with him in 1967, I sensed a flagging interest in the coordinate-free area, perhaps because of his many other interests—historical topics, measures of association, university administration and governmental statistics. In retrospect, his description of himself as an "overperfectionist" may provide some insight into the lack of a book based on his notes. The statistics community is certainly poorer because of this.

*Morris L. Eaton is Professor Emeritus, School of Statistics, University of Minnesota, Minneapolis, Minnesota 55455, USA e-mail: eaton@stat.umn.edu.*

My interest in the coordinate-free approach to linear statistical problems was motivated by at least two things: first, a predilection for elegant mathematics applied to statistics and second, the hint that such an approach could beneficially be brought to bear on multivariate analysis. It was with some trepidation that I approached Bill in early 1967 with a request to teach "his course" in the 1967–1968 academic year. He was thrilled that a young colleague had taken an interest in the coordinate-free methods and most likely, although unspoken, was pleased to get a break from the course. After teaching the course for two years running, I had a fairly firm grip on the material.

Before describing the influence of the coordinate-free approach on other research areas, let me briefly summarize the notes and Bill's approach. The initial chapter was an extended review of finite-dimensional inner product spaces. Paul Halmos's marvelous treatment of finite-dimensional vector theory is the obvious origin of this chapter, with full attribution of course. [A preliminary edition of the Halmos text, *Finite-Dimensional Vector Spaces*, was first copyrighted in 1942 by Princeton University Press. The text was published by Van Nostrand in 1958 and is currently in the Undergraduate Texts in Mathematics series published by Springer-Verlag (Halmos, 1974). Halmos was also at the University of Chicago during part of the 1950s and, with Jimmie Savage, had published the famous Halmos–Savage theorem in 1949 (Halmos and Savage, 1949).] After the vector space review followed the introduction of random vectors, mean vectors, covariance operators and the normal distribution. Linear model material including the Gauss–Markov theorem, hypothesis testing, confidence intervals and analysis of variance (ANOVA) examples rounded out most of the remaining chapters.

An early description of the Gauss–Markov theorem in a coordinate-free setting occurs in Kruskal (1961). This paper contains, in rather condensed form, a variety of the topics covered in the 1966 version of the lecture notes. Even in hindsight, it is difficult for me to assess the enormous influence Bill's notes had on both my mathematical skills and my research development. However, the direct effect of Kruskal (1968), a marvelous paper, is relatively easy to describe. In coordinate-free language, here is a statement of the main result of that paper:

> The Gauss–Markov and least squares estimators are the same if and only if the linear manifold of the mean vector is an invariant subspace of the covariance.

My first foray into coordinate-free multivariate analysis consisted of adapting the above result to a variety of multivariate analysis of variance (MANOVA) models. The material in Eaton (1970) attests to the power and insight provided by the Kruskal approach to linear models. Emboldened by this initial success, I set out to develop and extend multivariate analysis via a coupling of coordinate-free methods and invariance (group theoretic) arguments. The result of this work appeared in *Multivariate Statistics*: *A Vector Space Approach* (Eaton, 1983) that was published in 1983. The influence of the Kruskal notes on this work is palpable. Roughly half of the material represents multivariate versions of univariate results learned from teaching the coordinate-free course at Chicago. The importance of Bill's work in this application of beautiful mathematics to statistical theory is transparent.

Now let me turn to another aspect of the legacy of the coordinate-free approach. The School of Statistics at the University of Minnesota was founded in 1971. The adoption of a Chicago-like course into the school's curriculum was immediate and was most likely first taught by Kinley Larntz (Kinley's Ph.D. in 1971 was from Chicago). Other Minnesota faculty with interests in linear models (including Dennis Cook, Frank Martin and Sandy Weisberg) wholeheartedly adopted this approach and even today it is a mainstay in the Ph.D. curriculum at Minnesota. In addition and not surprisingly, Minnesota courses in multivariate analysis often have a coordinate-free flavor.

The standard setting of the Kruskal approach to linear models is a finite-dimensional inner product space. One consequence of choosing this framework is that the inner product determines the geometry of the vector space rather than the geometry being imposed by statistical considerations (such as the covariance structure of the linear model). In some linear model problems it is advantageous to use the "statistical geometry" of the problem rather than that dictated by a chosen inner product. For an example of such a problem, see Eaton (1985), where the full generality of Halmos's vector space development is brought to bear. This work represents a second generation of coordinate-free methods that are obviously motivated by Bill's contributions.

Influences of the coordinate-free approach on other writers can be found for example, in Arnold (1979, 1981) and Christensen (2002). The paucity of published works by Kruskal in this area makes it problematic to trace his impact in the development of linear model theory and methodology. His work of more than 40 years ago, however, was certainly a harbinger of our deeper understanding of this mainstream area of statistics.